\documentclass{iau}

\usepackage{amsmath}
\usepackage{graphicx}
\usepackage{multirow}

\begin{document}

\lefttitle{S. L. Yardley}
\righttitle{Solar Orbiter and Parker Solar Probe: Multi-viewpoint messengers of the inner heliosphere}

\jnlPage{1}{7}
\jnlDoiYr{2024}
\doival{10.1017/xxxxx}

\aopheadtitle{Proceedings IAU Symposium 390}
\editors{M. Romoli, L. Feng, M. Snow, eds.}

\title{Solar Orbiter and Parker Solar Probe: Multi-viewpoint messengers of the inner heliosphere}

\author{Stephanie L. Yardley}
\affiliation{Department of Mathematics, Physics and Electrical Engineering, Northumbria University, Newcastle Upon Tyne, NE1 8ST, UK}

\begin{abstract}
NASA's Parker Solar Probe and ESA/NASA's Solar Orbiter are encounter missions that are currently both in their nominal science phases, venturing closer to the Sun than ever before. These complementary spacecraft are operating together in order to combine in situ measurements of solar wind plasma in the inner heliosphere with high-resolution remote sensing observations of their source regions in the solar atmosphere. This paper highlights the synergetic science that these multi-viewpoint messengers of the inner heliosphere enable and how they are working together to significantly advance our understanding of the physical processes that are important for solar wind formation, the eruption of coronal mass ejections and their space weather effects.
\end{abstract}


\maketitle

\section{Near-Sun Multi-viewpoint Messengers}

\begin{figure}[t]
\begin{center}
 \includegraphics[width=5in]{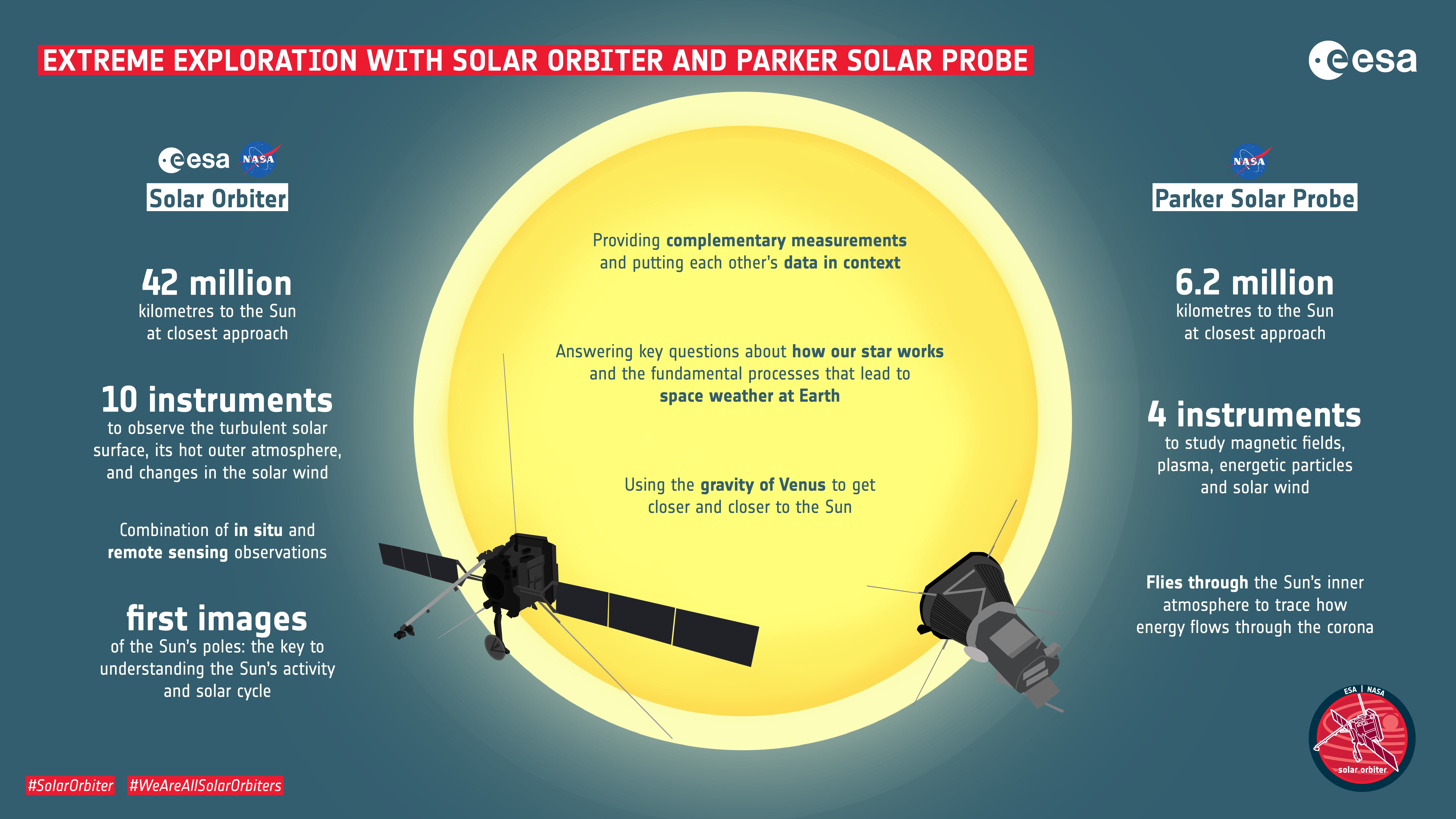} 
\caption{PSP and SO are multi-viewpoint messengers of the inner heliosphere working together to understand the complexities of our star. Image courtesy of ESA/S.Poletti \url{https://www.esa.int/ESA_Multimedia/Images/2020/01/Extreme_exploration_with_Solar_Orbiter_and_Parker_Solar_Probe}}
   \label{fig:fig1}
\end{center}
\end{figure}

Many of the fundamental processes responsible for the formation and complex structure of the heliosphere occur at close heliocentric distances meaning that by the time solar wind and solar transients arrive near-Earth at 1~au, their structures have evolved and information regarding their origin has been lost. However, we are now entering a golden era of heliospheric physics due to the existence of the first orbiter missions that have the ability to capture in situ measurements of plasma and magnetic fields in the inner heliosphere and connect them to remote-sensing observations of their source regions in the Sun's atmosphere. NASA's Parker Solar Probe (PSP; \citealt{fox2016}) and ESA/NASA's Solar Orbiter (SO; \citealt{muller2020, garcia2021}) are currently venturing closer to the Sun than ever before as they are both in their nominal mission phases. These missions are working together, providing complementary observations and measurements, in order to understand how the Sun creates and controls the heliosphere and the activity within (Figure~\ref{fig:fig1}). 

PSP launched on board a Delta IV Heavy rocket in August 2018 from Cape Canaveral and consists of a suite of four instruments including the FIELDS experiment \citep{bale2016}, the Integrated Science Investigation of the Sun (IS$\odot$IS; \citealt{mccomas2016}), the Wide Field Imager for Solar Probe (WISPR;\citealt{vourlidas2016}), and the Solar Wind Electrons Alphas and Protons (SWEAP; \citealt{kasper2016}). This suite of instruments operates together to: understand the heating and acceleration of the corona and solar wind; uncover the structure and dynamics of the coronal magnetic field; and understand the physical processes involved in accelerating energetic particles by orbiting the Sun closer than any spacecraft before it. Due to its close proximity, this allows for periods of co-rotation where PSP will sit above the same source region of the solar wind, allowing for the analysis of the evolution and small-scale variability within the same stream.

PSP encounters are defined by PSP coming within a distance of less than 0.25~au from the Sun, with PSP having just completed its 21st encounter on 30th September 2024 reaching a perihelion of 0.048~au (10.4~R$_{\odot}$), while moving at a record-breaking 394,000~mph. The mission is currently in its nominal phase but will make one final Venus gravity assist maneuver on 6 November 2024 before making the first of its three final closest approaches to the Sun beginning 24 December 2024 (with subsequent approaches on March 2025 and June 2025). This will allow the spacecraft to come within just 6.2 million km or 0.04~au (8.79~R$_{\odot}$) while moving at 430,000~mph. The nominal mission is 7 years in total, but PSP has fuel onboard for a potential extended mission phase, where the distance will remain the same for any subsequent perihelia occurring during the potential extended mission phase \citep{velli2020, raouafi2023a}.

SO launched on an Atlas V 411 rocket also from Cape Canaveral, less than two years later in February 2020, becoming the most complex spacecraft to be sent close to our Sun. The spacecraft has six remote-sensing instruments \citep{auchere2020} including: the Extreme Ultraviolet Imager (EUI; \citealt{rochus2020}), Metis \citep{antonucci2020}, Solar Orbiter Heliospheric Imager (SoloHI; \citealt{howard2020}), Spectral Imaging of the Coronal Environment (SPICE; \citealt{spice2020}), Spectrometer/Telescope for Imaging X-rays (STIX; \citealt{krucker2020}), Polarimetric and Helioseismic Imager (PHI; \citealt{solanki2020}, and four in situ instruments \citep{walsh2020} including: the Energetic Particle Detector (EPD; \citealt{pacheco2020})), the Magnetometer (MAG; \citealt{horbury2020}), Radio and Plasma Waves (RPW; \citealt{maksimovic2020}) and the Solar Wind Analyser (SWA; \citealt{owen2020}). The payload of 10 instruments observes jointly to: understand the origins of the solar wind and the coronal magnetic field; how variability is driven by solar transients; how solar eruptions produce energetic particle radiation that fills the heliosphere; and how the solar dynamo operates and drives the Sun-heliospheric connection. Periods of near-corotation, not only allow SO to observe the same solar wind stream for several days but to link the small-scale variability within the stream at the spacecraft back to the processes and evolution occurring in the source region on the Sun. These periods also provide opportunities to capture the long-term evolution and the physical processes that occur in the build-up to the eruption of active regions, important for advancing space weather forecasting. 

In order to achieve these top-level science goals, the science activities of SO must be planned in advance. This is due to the unique orbit and telemetry constraints of the spacecraft, the spatial FOV of the high-resolution remote-sensing instruments, and the three 10-day remote-sensing windows (RSWs) centered around perihelion per orbit. Therefore, the science activities, which have been built on contributions from the international solar and heliospheric physics community, are carried out through Solar Orbiter Observing Plans (SOOPs; \citealt{zouganelis2020}), where the remote-sensing observations and in situ measurements are coordinated. This coordination takes place on long and very short-time scales, from deciding the observational modes of the instruments for 6 month periods to target selection approximately three days prior to the observations i.e. pointing of the spacecraft and the remote-sensing instruments at specific features on the Sun. Not only does the long-term planning ensure that SO achieves its top science goals but it also allows close coordination between SO and PSP during favourable orbital configurations (see Table~1 and Section 3.2 of \citealt{velli2020}). The SO-PSP Coordination Working Group\footnote{\url{https://s2e2.cosmos.esa.int/confluence/display/SOSP/SO-PSP+Coordination+WG)}} was established for this purpose. Advanced planning also allows for complementary observation campaigns to be run by external facilities including space-based assets such as Hinode \citep{kosugi2007} and IRIS \citep{depontieu2014} through IRIS Hinode Operation Plans (IHOPs), and ground-based telescopes such as Swedish Solar Telescope (SST; \citealt{scharmer2003}) and the Daniel K. Inouye Solar Telescope (DKIST; \citealt{rimmele2020}) to be coordinated. In addition, tools and techniques \citep{rouillard2020} have been developed by the ESA's Modelling and Data Analysis Working Group (MADAWG) in order to support the short-term planning and post-observation analysis of SO and other coordinated datasets.

SO entered its four-year nominal mission phase on 26 November 2021, with the first close perihelion passage during this phase occurring on 26 March 2022, reaching a distance of 0.32~au from the Sun. Subsequent perihelia then occur roughly every 6 months at heliocentric distances less than 0.3~au, in which the three 10-day RSWs are centered around. In February 2025, a Venus gravity assist maneuver will occur to increase the inclination of SO to 17$^{\circ}$ and bring SO at its closest distance from the Sun (42 million km or 0.28~au). Following flybys in December 2026 and beyond will mark the "high-latitude" or extended phase of the mission, where the inclination of SO will be increased to up to 33$^{\circ}$. This will allow Solar Orbiter to detect fast solar wind from the polar regions but also to take the first ever images of the small-scale magnetic features at the poles, which are key in order to understand the Sun's magnetic activity cycle.

While there are some differences in the science goals of PSP and SO, such as SO observing the poles during the extended phase of the mission, they are complementary spacecraft operating together to better understand how our Sun works and the physical processes that lead to space weather (Figure~\ref{fig:fig1}). This paper illustrates and presents examples of how the unique observing opportunities and viewpoints provided by PSP and SO, both jointly and separately can advance our understanding of the origins of the solar wind and solar eruptions.

\section{The near-Sun Solar Wind}

\subsection{Results from PSP early nominal and SO cruise phase}

The first few years of PSP encounters occurred during solar minimum, meaning that mainly slow solar wind streams were observed and there were few major solar eruptive events \citep[for an indepth review of these results see][]{raouafi2023a}. However, this allowed for the dynamic and variable nature of newly formed solar wind to be uncovered. The first perihelion encounter of PSP took place in November 2018, where PSP was magnetically connected \citep{bale2019, badman2020} to a small equatorial coronal hole with predominantly negative polarity, during the two-week co-rotation period centered around perihelion, from which mostly Alfv{\'e}nic slow solar wind, \citep[for a review on this topic see][]{damicis2021}, was detected. During this period, measurements taken by the FIELDS and SWEAP instruments revealed sudden polarity reversals of the radial magnetic field, opposite to the dominant magnetic field polarity \citep{bale2019}. Further supporting evidence that these are reversals rather than changes of magnetic connectivity across solar source regions are provided by the electron pitch angle distributions. The period of these field reversals, known as ``switchbacks", were found to occur on timescales ranging from seconds, minutes to even longer reversals that lasted upwards of an hour or more \citep{dudok2020}. These reversals are highly correlated with large-amplitude spikes or fluctuations in the radial solar wind velocity \citep{kasper2019}, corresponding to outwardly propagating large-amplitude Alfv{\'e}n waves. Similar structures have been sporadically detected previously at 0.3~au by Helios \citep{borovsky2016, horbury2018}, at 1~au by ACE \citep{gosling2009,owens2013}, and beyond 1~au during polar passes made by Ulysses \citep{balogh1999, yamauchi2004}. But what is surprising about these reversals is their ubiquitous appearance, large-amplitudes, and deflection in the tangential field direction \citep{kasper2019, dudok2020}, along with the fact that they were observed in mostly Alfv{\'e}nic slow solar wind.

\begin{figure}[t]
\begin{center}
 \includegraphics[width=5in]{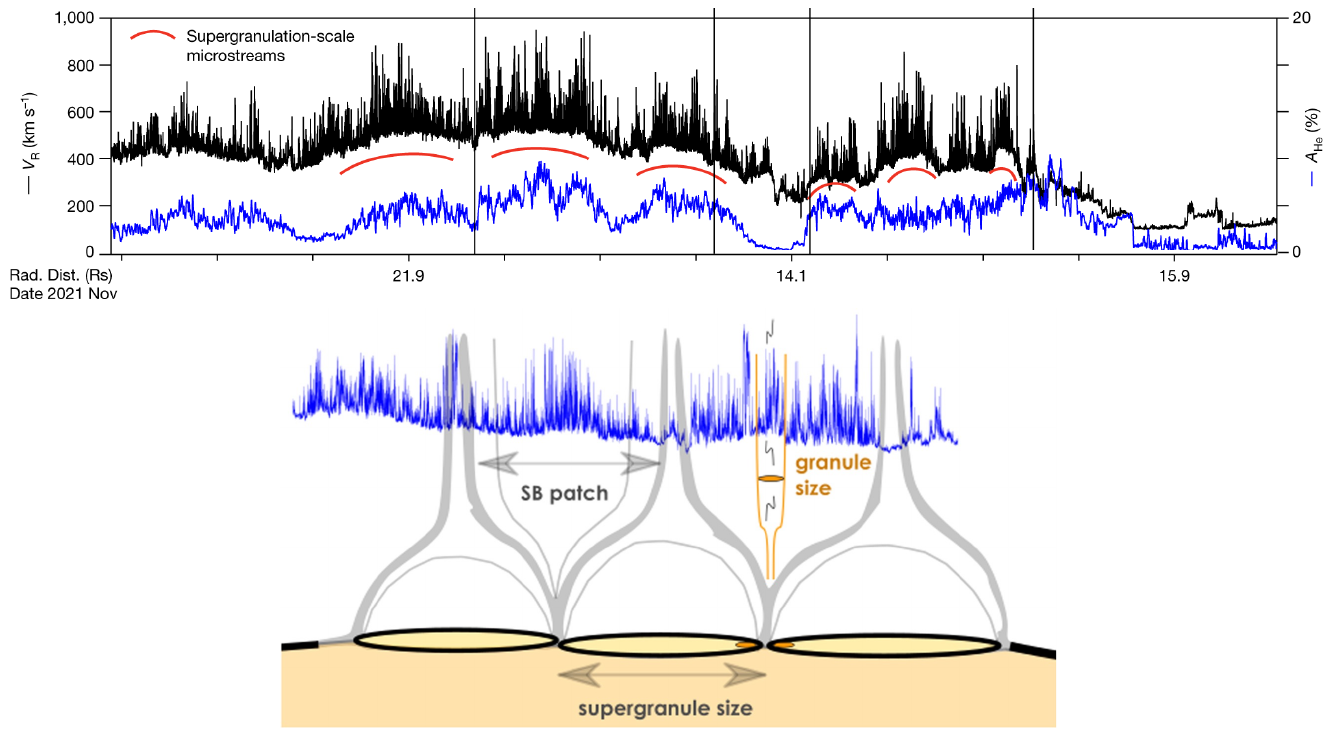} 
\caption{The microstream structure of switchbacks and their postulated origins. Top panel: The radial velocity of the solar wind and the thermal alpha particle abundance as measured by the SWEAP instrument during the period of 20-21 November 2021 during PSP encounter 10 \citep[Figure adapted from][]{bale2023}. The microstream structure of the velocity spikes are shown by the arcs. Bottom panel: An illustration of how the microstream structure of switchbacks is modulated by granular and supergranular cells. The radial magnetic field measurements taken by FIELDS during encounter 5 is plotted. Thick lines indicate regions of open-closed magnetic fields at the supergranular cell boundaries \citep[Figure adapted from][]{fargette2021}.}
   \label{fig:PSPswitchbacks}
\end{center}
\end{figure}

The exact origins of these switchbacks are heavily debated, and as they could provide insights into the origins of the slow solar wind \citep[see reviews by][]{abbo2016,cranmer2017,viall2020}, the heating of the solar corona, and the acceleration of the solar wind, there is a huge motivation to understand these structures. Measurements during multiple PSP encounters suggest that switchbacks occur in patches or microstreams (Figure~\ref{fig:PSPswitchbacks}), with a spatial scale that is consistent with magnetic funnels or plumes in the corona, associated with the boundaries of supergranulation cells \citep{bale2021, fargette2021, bale2023}. The asymmetric nature of the microstreams along with their association with an increased particle abundance suggests that these structures are generated by interchange reconnection between open and closed magnetic fields in the corona \citep{fisk2005, fisk2020}. More recent work \citep{hou2024} has tied the in situ measurements of switchbacks made by PSP to SDO observations of jets at chromspheric network boundaries mostly associated with magnetic flux cancellation rather than emergence although, this relation should be explored further, particularly with high-resolution observations of jets taken by SO/EUI.

However, there have been various mechanisms proposed to generate switchbacks, which can generally be split into processes taking place in the solar atmosphere (reconnection or loop opening processes) or locally in the solar wind (wave/turbulence driven processes). Proposed switchback generation mechamisms include: interchange reconnection creating kinked field lines, jets or flux ropes \citep{fisk2020, zank2020, sterling2020, drake2021}, velocity shear and footpoint motions \citep{ruffolo2020, schwadron2021}, and expanding Alfv{\'e}n waves and turbulence \citep{squire2020, mallet2021, shoda2021}.

Due to the diversity of mechanisms proposed for switchback generation, it is clear that multi-viewpoint measurements along with high-resolution remote-sensing observation of their potential coronal sources are required. The first opportunity for joint PSP-SO observations came during the 6th encounter of PSP and SO's cruise phase in September 2020, when PSP-SO conjunction occurred on 27th September 2020. PSP (0.09~au) and SO (0.98~au) were magnetically connected to the same coronal hole \citep{horbury2021, federov2021} as determined by the magnetic connectivity tool\footnote{\url{http://connect-tool.irap.omp.eu/}}. This means that PSP and SO both observed solar wind ejected from the same source during similar but not exactly the same time intervals. SO detected short magnetic field fluctuations and reversals similar to the switchbacks measured by PSP. These studies proposed that the electron pitch angle distributions and alpha-particle speed distributions suggest that these fluctuations are formed due to solar wind velocity shear similar to \citet{schwadron2021} however, further out reconnection occurs between the folded field line leads to the formation of a flux rope later observed by SO.

\begin{figure}[t]
\begin{center}
 \includegraphics[width=5in]{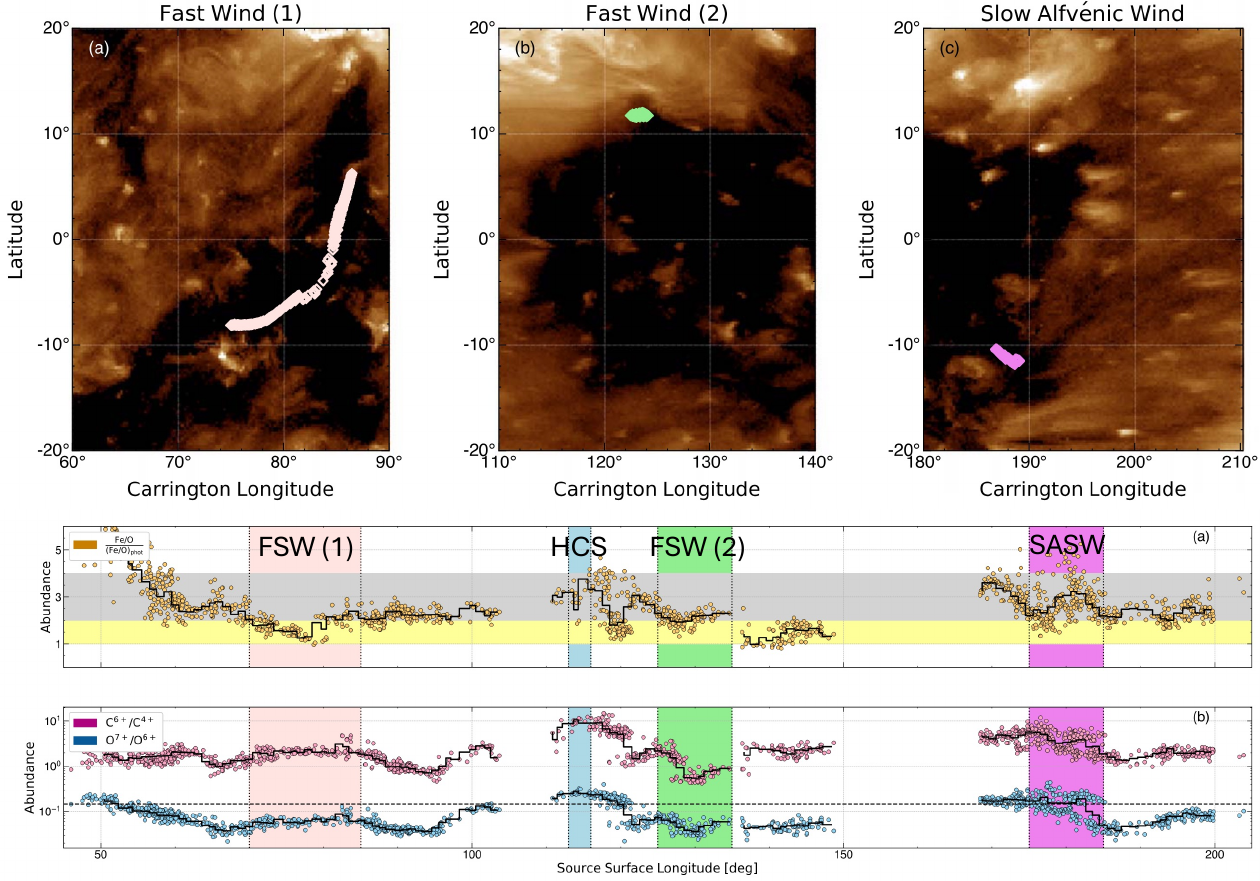} 
\caption{Observations taken during PSP-SO conjunction during PSP encounter 11. Top panel: SDO/AIA images of the source regions of the two fast and one Alfv{\'e}nic slow solar wind streams highlighted in the bottom panel. The diamonds represent the connectivity footpoints of PSP. Bottom panel (a): The Fe/O ratio as measured by SO SWA/HIS, where the values have been normalised \citep{asplund2021}. The shaded regions correspond to where the Fe/O ratio is between 1-2 and 2-4. (b) Charge state ratios measured by SO SWA/HIS. The black dashed line represents the threshold of 0.145 for streamer wind \citep{wang2016}. Both panels are plotted as a function of source surface longitude where the shaded regions correspond to wind streams from the sources in the top panel apart from HCS which is the heliospheric current sheet crossing. Figure adapted from \citet{ervin2024}.
} \label{fig:SO_PSP_Wind}
\end{center}
\end{figure}

In February 2022, just prior to the start of SO's first close perihelion passage during the nominal phase of the mission, and during PSP encounter 11, PSP and SO were again in conjunction. \citet{ervin2024} utilized this opportunity to comprehend the variability and composition of fast and Alfv{\'e}nic slow streams emerging from multiple coronal holes. The time periods where the spacecrafts were radially aligned measuring the same wind streams were identified along with their sources by using both potential field and MHD modelling. It was found that the two models were in generally good agreement. The two fast wind streams that originated from two large coronal holes are associated with low charge state and Fe/O ratios as measured by the SWA/Heavy Ion Sensor (HIS) suggesting photospheric abundances and temperatures, as expected (Figure~\ref{fig:SO_PSP_Wind}). However, the Alfv{\'e}nic slow solar wind showed distinct populations of plasma with varying characteristics, including a region with high value charge state and Fe/O ratios suggesting coronal abundances and another with low charge state ratios and a low Fe/O ratio typical of photospheric abundances. A variance is also observed in the alpha-to-proton abundance. These results suggest that the Alfv{\'e}nic slow solar wind originates from an over-expanded boundary of a coronal hole.

Another study by \citet{rivera2024} determined that PSP and SO crossed the same fast solar wind stream from the same source in the corona within 2 days of each other on 25 February 2022 (13.3$R{\odot}$) and 27 February 2022 (127.7~R$_{\odot}$), respectively. The two spacecraft crossed the streams in opposite directions, with PSP crossing the same source much faster. However, given a similar transit time (~40-45~hrs) of the solar wind plasma to arrive at each spacecraft it allows the comparison between their in situ measurements, which support this conclusion. By using these measurements and quantifying the different energy contributions in the expanding solar wind, they show that switchbacks and large amplitude Alfv{\'e}n wave provide the energy required to heat and accelerate the solar wind. 

\subsection{Results from SO's first close perihelion passage during nominal phase}

The first close perihelion passage of SO took place with three RSWs occurring between 2 March 2022 and 6 April 2022, with a heliocentric distance ranging from 0.32 to 0.55 au, allowing SO to use all ten instruments to take the closest remote-sensing observations of the Sun and link them to in situ measurements of the solar wind. A description of the Solar Orbiter Observing Plans (SOOPs) that ran during this period for this purpose can be found in both \citet{zouganelis2020} and Section 2.3 of \citet{berghmans2023}. Information on SOOP operation, coordinators, and supporting observations from external facilities can be found on the ESA SOOP summary page\footnote{\url{https://www.cosmos.esa.int/web/solar-orbiter/soops-summary}}. In this section, we will describe the results from these particular SOOPs focused on connecting in situ measurements of solar wind to remote-sensing observations of their sources. 

At the very beginning of the first RSW on 2 March 2022 at around 0.54~au, SO ran the first instance of the Composition Mosaic SOOP (L\_SMALL\_MRES\_MCAD\_ConnectionMosaic). This SOOP was led by the SPICE instrument, which performed three raster scans of a portion of the Sun from North to South, in order to optimize the chance of catching the region on the Sun that is magnetically connected to Solar Orbiter. One of the main diagnostics used to trace plasma detected in situ back to its solar source is plasma composition or elemental abundances. This relies on the First Ionisation Potential (FIP) effect \citep{laming2017}, the enhancement of low-FIP elements in the corona. In this respect, SPICE can provide composition diagnostics that can be linked with measurements taken by SWA/HIS to be able to connect solar wind plasma measured in situ to its origins in the solar corona. 

During the mosaic observations, SPICE caught two active regions in the three scans (see Figure), NOAA ARs 12957 and 12958, close to a large equatorial coronal hole \citep{varesano2024}. In order to determine the plasma composition diagnostics of the regions they utilised and compared three different methods based on spectral line ratios and differential emission measure inversion. \citet{varesano2024} found enhanced plasma composition at the base of the coronal loops associated with the two active regions. The magnetic connectivity footpoints of SO, determined by using the magnetic connectivity tool, were situated at the boundary of the emerging AR 12957 for the corresponding in situ period (3 to 5 March 2022). This connectivity was preliminarily verified through the Fe/O ratio, which showed an increasing enhancement throughout the time interval, indicating that AR 12957 could be the source of slow solar wind later detected in situ by SO.

\begin{figure}[t]
\begin{center}
 \includegraphics[width=5in]{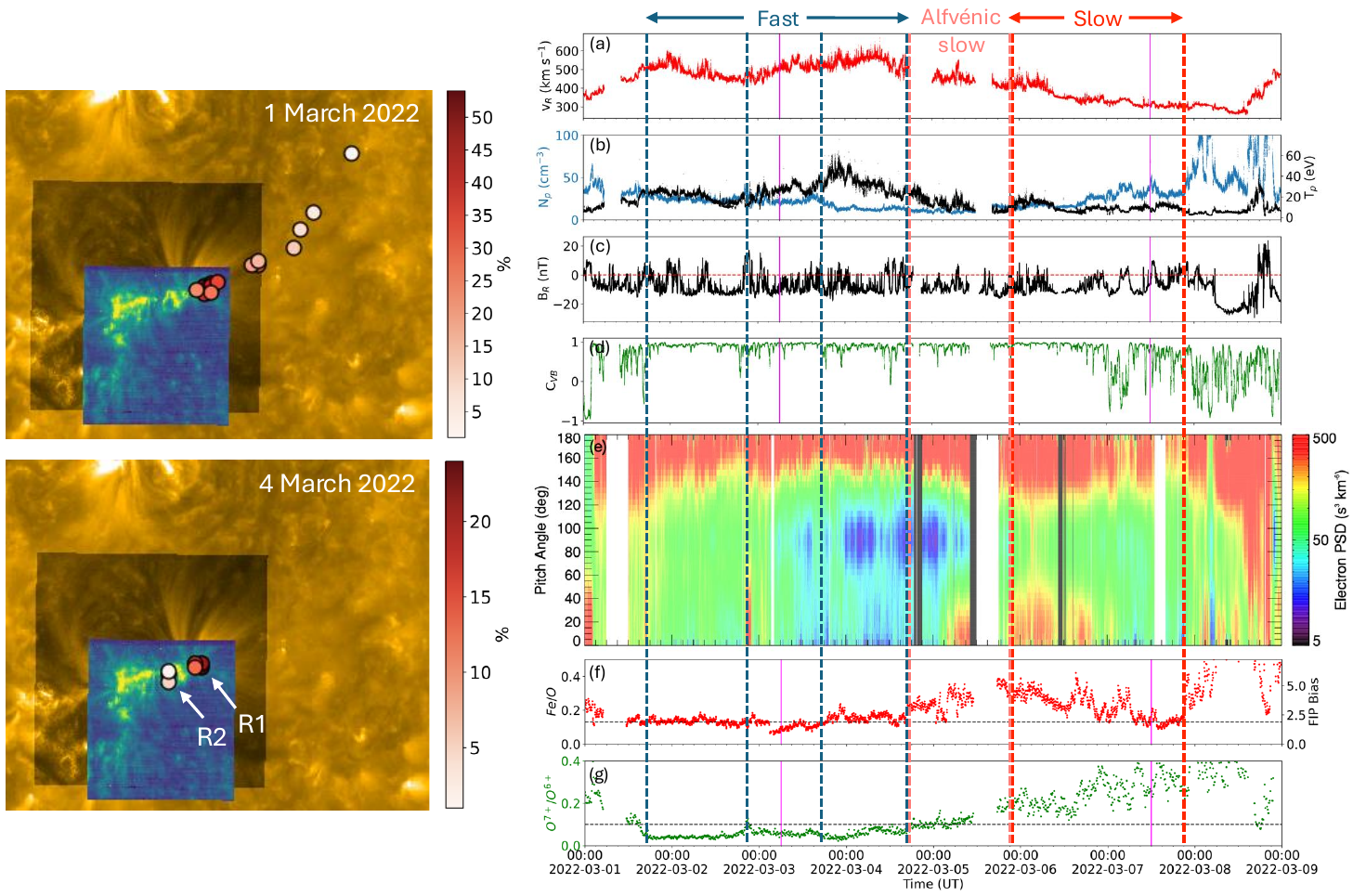} 
\caption{Remote-sensing observations and in situ measurements taken at the time of the Slow Wind Connection Science SOOP during SO's first perihelion. Left panel: The connectivity footpoints of SO taken from the magnetic connectivity tool are overlayed on a composite image of EUI/FSI, EUI/HRI and SPICE corresponding to the Sun times of 1 and 4 March 2022. The labels R1 and R2 correspond to the leading polarities of ARs 12961 and 12957, respectively. The gradient of the points show the percentage probability of the connectivity. Right panel: The in situ parameters measured by SWA/PAS (a,b), MAG (c), VB correlation factor showing Alfv{\'e}nic content (d), SWA/EAS (e), and SWA/HIS (f,g). The fast, Alfve{\'e}nic slow and slow wind streams are labelled and indicated by dashed lines. The two solid lines correspond to the in situ timings of the connectivity shown in the left panel. Figures courtesy of: \url{https://www.cosmos.esa.int/web/solar-orbiter/-/science-nugget-multi-source-connectivity-drives-heliospheric-solar-wind-variability} and adapted from \citet{yardley2024}. }
   \label{fig:slowwind}
\end{center}
\end{figure}

Immediately after the connection mosaic SOOP in RSW1, from 3 to 6 March 2022 (0.55-0.51~au), the Slow Solar Wind Connection Science (L\_SMALL\_HRES\_HCAD\_Slow-Wind-Connection) SOOP operated for the first out of two instances during this perihelion \citep{yardley2023}. The primary targets of this SOOP are open-closed magnetic field boundaries such as that at the edges of active regions and/or coronal holes, potential sources of slow solar wind. Again, selecting the observational target for this SOOP and the post-observational analysis relied upon the magnetic connectivity tool. For the duration of the first instance of the Slow Wind Connection Science SOOP, SO targeted the boundary of AR 12957 that was also observed during the connection mosaic. During the time period, there was continuous flux emergence occurring in this region leading to the formation of an active region complex \citep{yardley2024}. Observations by EUI's High Resolution Imager (HRI), along with a potential field extrapolation showed that the region exhibited large-scale coronal fan loops associated with open magnetic fields. The post-observed connectivity analysis, provided by the magnetic connectivity tool, show that the connectivity footpoints of SO transition across a dark channel in EUV, adjacent to the large equatorial coronal hole to the two leading polarities of the active region complex (Figure~\ref{fig:slowwind}). This is the same equatorial coronal hole that PSP was connected to a few days earlier (top left panel of Figure~\ref{fig:SO_PSP_Wind}). The composition analysis carried out on the spectroscopic data from SPICE revealed that there was a spatial difference in the composition (from coronal to photospheric) in the two leading negative polarities (R1 and R2 of Figure~\ref{fig:slowwind}) of the active region complex.

It was apparent that multiple solar wind streams with varying characteristics were detected by SO. This included fast streams from the large equatorial coronal hole to Alfv{\'e}nic slow to typical non-Alfv{\'e}nic slow solar wind from the two regions of the active region complex. This scenario is supported by the solar wind plasma properties including the decreasing solar wind speed measured by SWA/Proton and Alpha particle Sensor (PAS), the negative radial magnetic field direction measured by MAG, the electron strahl population measured by SWA/Electron Analyser System, the Alfv{\'e}nic content of the wind accompanied by episodes of switchbacks, and the increase in heavy ion charge state and Fe/O ratios measured by SWA/HIS (right panel of Figure~\ref{fig:slowwind} and \citet{yardley2024}). The continuous flux emergence occurring in the central part of the active region complex along with the symmetrical bidirectional electron strahl and the arrival of highly Alfv{\'e}nic slow solar wind favour solar wind plasma escaping from the core of the active region complex via the process of interchange reconnection. Furthermore, these results show that the variability of the solar wind is driven by the changing connectivity across the multiple source regions and the change in topology of the source regions themselves.

Another instance of the Slow Solar Wind Connection Science SOOP ran between 17 March 2022 and 22 March 2022 (0.38-0.34~au), during RSW2. In this instance, two observational targets were selected, including the boundary of the southern polar coronal hole and a small-decayed positive polarity of an active region in the northern hemisphere, due to the drastically changing magnetic connectivity of Solar Orbiter during this period. On 18 March 2022, SO captured the rare occurrence of a coronal dimming associated with a filament eruption merging with the southern polar coronal hole \citep{ngampoopun2023}, which was also observed by Hinode and IRIS \citep{yardley2023}, allowing the first spectrosopic analysis of this type of interaction. Unusually, the merged region was found to persist in the EUI and SDO observations for at least 72 hours with the upflow velocities measured by Hinode/Extreme-ultraviolet Imaging Spectrometer (EIS) in the coronal dimming becoming comparable to the coronal hole after the merging. The study suggested that component reconnection plays an important role in the merging of these structures, and the persistence of the resulting merged structure.

The remaining three days of the second instance of the Slow Solar Wind Connection Science SOOP was spent observing the positive polarity of AR 12967. Tracking this region with SO/EUI, SDO and Hinode/EIS revealed a narrow dark corridor in EUV where strong plasma upflows were observed, associated with open magnetic field as evident from a potential field extrapolation. The magnetic field extrapolation along with maps of the squashing factor Q, which show drastic changes in the field line mapping, demonstrated that this narrow corridor super-radially expands by about 60${^\circ}$ latitude by the source surface at 2.5~$R_{\odot}$. The narrow open-field corridor was determined to be the source region of a slow solar wind stream associated with moderate Alfv{\'e}nicity and switchbacks. The in situ measurements detected by SO show extremely low solar wind speeds and proton temperatures and high proton densities. These results provide supporting evidence of the S-web theory \citep{antiochos2011}, where slow solar wind originates from a network of narrow open-field corridors in the corona via interchange reconnection. It is also plausible that reconnection events occurring at the narrow open-field corridors, that form part of the S-web, could be responsible for the generation of switchbacks. 

\begin{figure}[t]
\begin{center}
 \includegraphics[width=4.5in]{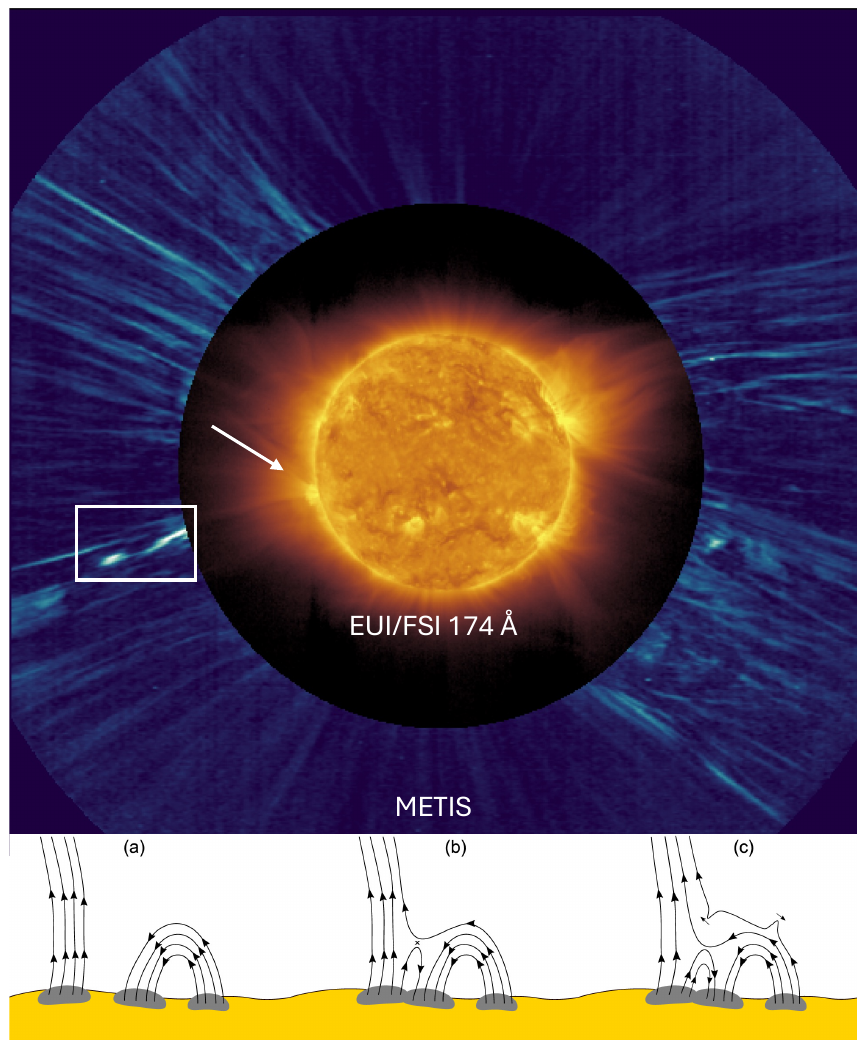} 
\caption{Remote-sensing observations taken during SO's first perihelion passage and the CH Boundary Expansion SOOP on 25 March 2022, showing the S-shaped structure identified as a switchback and the proposed mechanism behind its formation. Top panel: Composite image of EUI/FSI 174~\AA\ and Metis total brightness (tB) adapted from \citep{telloni2022}, that has been processed to bring out small-scale dynamic structures using a Simple Radial Gradient Filter (SiRGraF) algorithm \citep{patel2022}. Bottom: Schematic of proposed mechanism through interchange reconnection at closed-open magnetic field boundaries of an active region. Image courtesy of ESA: \url{https://www.esa.int/ESA_Multimedia/Images/2022/08/Creating_a_solar_switchback}  
} \label{fig:SO_switchback}
\end{center}
\end{figure}

The final SOOP to operate during RSW2 of the first perihelion passage on this topic was the Coronal Hole Boundary Expansion (L\_BOTH\_HRES\_LCAD\_CH-Boundary-Expansion) SOOP. The SOOP ran between between 25 March 2022 and 27 March 2022 with the aim to study coronal hole boundaries as the sources of the slow solar wind using Metis. During this SOOP, at a distance from the Sun of 0.32~au, Metis imaged a single, large propagating S-shaped structure (Figure~\ref{fig:SO_switchback}) interpreted as the first observed switchback in the solar corona \citep{telloni2022}. The S-shaped structure originated above the closed loops of an active region visible in EUI, a region which is surrounded by coronal holes associated with open magnetic fields as derived from extrapolated magnetic field lines from a 3D MHD model developed by Predictive Science Incorporated (PSI). They suggest that this S-shaped structure interpreted as a switchback was created through interchange reconnection high up in the corona between the large-scale closed active region loops and surrounding open magnetic fields. Particularly in the region of a null point evident by high values in the squashing factor Q synoptic maps. 

Since the first perihelion passage of SO, the SOOPs detailed above with the addition of the Fast Wind (L\_SMALL\_HRES\_HCAD\_Fast-Wind) SOOP, have operated multiple times during subsequent perihelia. The Fast Wind is  similar to the Slow Solar Wind Connection Science SOOP but has been optimised in order to measure fast winds originating from the centers of coronal holes. We can expect more results from these SOOPs imminently and also joint observations from PSP-SO due to an increasing amount of alignments.  Of particular interest will be some of the slow and fast wind SOOP observations that have been taken during SO's perihelia in 2023 and 2024 that correspond to some of the later PSP encounters. This includes observations where PSP and SO are both at perihelion (0.1 and 0.3~au) and are radially aligned.

\section{Eruption and Propagation of CMEs in the inner heliosphere}

Since the launch of PSP, many coronal mass ejections (CMEs) and their interplanetary counterparts (ICMEs) have been observed however, not necessarily simultaneously with remote-sensing observations and in situ measurements. There have been cases where CMEs have been observed remotely by WISPR \citep{howard2019, hess2020, liewer2020, wood2020, braga2021, howard2022} however, the ICMEs were not detected in situ. On the other hand, in situ measurements have been obtained for ICMEs, some of which were not visible from L1, were slow stealth CMEs i.e. not associated with low coronal eruption signatures \citep{nitta2021}, or occurred during time periods where PSP was at further distances from the Sun. A description of the studies of some of these events, occurring during the first six encounters of PSP, are given in \citet{raouafi2023a}. A list of 28 ICMEs detected in situ by PSP during October 2018 and August 2022 has also been constructed by \citet{salman2024}.
There have also been multiple CMEs observed by Metis \citep{andretta2021} and/or SoloHI \cite{bemporad2022} along with multiple spacecraft \citep{andretta2021, hess2023}. A useful multi-spacecraft CME catalogue called LineupCAT\footnote{\url{https://www.helioforecast.space/lineups}} is under construction by HELIO4CAST project, listing ICMEs observed by multiple spacecraft including but not limited to PSP and SO \citep{mostl2022}.

In this section, we focus only on studies that involve ICMEs detected along the Sun-Earth line or that are observed by both PSP and SO in order to tackle the issue of predicting the arrival and magnetic field orientation of flux ropes embedded in CMEs for geoffectiveness and space weather prediction purposes \citep{owens2020}.

\subsection{Results from PSP and SO nominal phase}

\begin{figure}[t]
\begin{center}
 \includegraphics[width=5in]{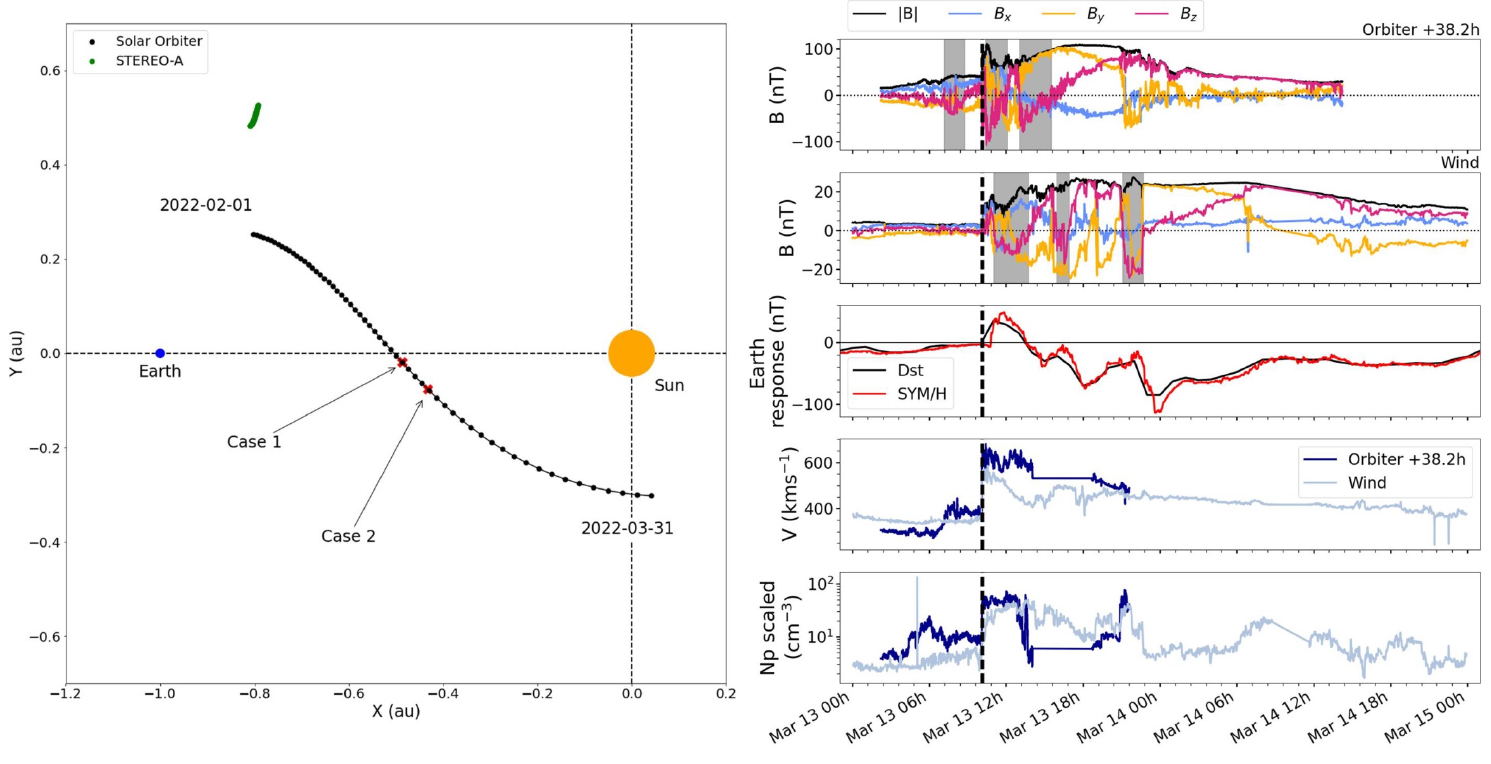} 
\caption{The position and in situ measurements taken by SO during the first perihelion passage. Left panel: SO's orbit in GSE coordinates between 1 February 2022 and 31 March 2022. Case 1 and Case 2 refer to the two CMEs that occurred on 7 March 2022 and 11 March 2022, just after SO crossed the Sun-Earth line at a distance of $\sim$0.5~au. Right panel: The in situ measurements for the second CME (Case 2). From top to bottom shows: the magnetic field components measured by MAG on board SO shifted by 38.2hr to align with the shock front (thick dashed line) measured by Wind at 1~au, the geomagnetic storm at Earth shown by the DST and SYM/H indices, the velocity and proton density as measured by SO and Wind, where the proton density measured by SO has been scaled by 1/r$^{2}$. The shaded regions indicate the three intervals of negative B$_{z}$. Figure adapted from \citet{laker2024}.
} \label{fig:SO_Bz}
\end{center}
\end{figure}

During SO's first perihelion passage in March 2022 just after it had crossed the Sun-Earth line it was possible to utilize SO as an upstream space weather forecasting monitor, given its capability at this position to return low latency data within a sufficient time frame in order to predict the arrival and geoeffectiveness of CMEs arriving at Earth \citep{laker2024}. At this time, SO was at approximately 0.5~au when it detected two CMEs on 7 and 11 March 2022, both of which were then later detected at L1 (left panel of Figure~\ref{fig:SO_Bz}). For these two events, it was possible to use the speed from SWA/PAS to predict the estimated arrival time of the CMEs at Earth. Even with a simple estimation it was possible to give accurate Earth arrival time, predictions within 1-3~hrs, for both CMEs. A more complex analysis was conducted using the ELEvoHI model
\citep{amerstorfer2021, bauer2021} and heliospheric imager data from STEREO-A, which allowed the mean absolute error in the arrival time to be reduced from 10.4 to 2.5 hrs and 2.7 to 1.1 hrs for the two CMEs. By constraining the model with data from 0.5~au it was therefore possible to update the prediction before the CMEs arrived at Earth, providing a minimum warning of $\sim$40~hrs. 

With the combination of data from WIND at 1~au and Mag onboard SO it was also possible to compare the magnetic structure of the arriving CMEs at different heliocentric distances (right panel of Figure~\ref{fig:SO_Bz}). For the second CME there was excellent agreement between the two spacecraft for the z-component of the magnetic field (B$_{z}$). They both detected three time periods of negative B$_{z}$ followed by a northward orientated flux rope. The magnetic response at Earth was extremely well correlated to the magnetic configuration of the flux rope that was observed by SO at 0.5~au around 40~hrs prior. However, the in situ measurements of the first CME were more complicated with the interaction and arrival of two flux rope structures. While this kind of analysis is promising in terms of using spacecraft such as SO as a real-time space weather forecasting monitor, the analysis of a larger sample of events is needed, particularly as flux rope structures and orientation can significantly evolve on their journey to 1~au from the Sun.

During the final RSW of the first close perihelion passage, SO was operating the long-term active region (R\_SMALL\_MRES\_MCAD\_AR-Long-Term) SOOP between 31 March and 4 April 2022. The goal of this SOOP was to monitor an active region beyond its disk passage, particularly during the decay phase when filaments form and later erupt at CMEs. During the observation period, SO targeted two decaying active regions at this time: ARs 12975 and 12976, between which a filament had formed \citep{janvier2023}. During the observations on 2 April, two M-class flares occurred, the first of which was associated with a filament eruption. Given the position of SO with respect to Earth, this allowed for the first time a stereoscopic spectroscopic view of an eruption. The eruption of the filament was triggered by the emergence of a small parasitic bipole underneath the filament and between the two active regions as observed by PHI/HRT. The subsequent reconnection led to not only the triggering of the eruption but also the formation of multiple loop and flare ribbon systems as seen by EUI and SPICE on board SO, as well as from IRIS and Hinode. It was determined that the eruption was consistent with the standard 3D flare model but this interpretation was only possible due to the range of viewing angles and spectroscopic instruments that observed the eruption. The resulting ICME was also detected in situ for which the analysis is ongoing.

\begin{figure}[t]
\begin{center}
 \includegraphics[width=5in]{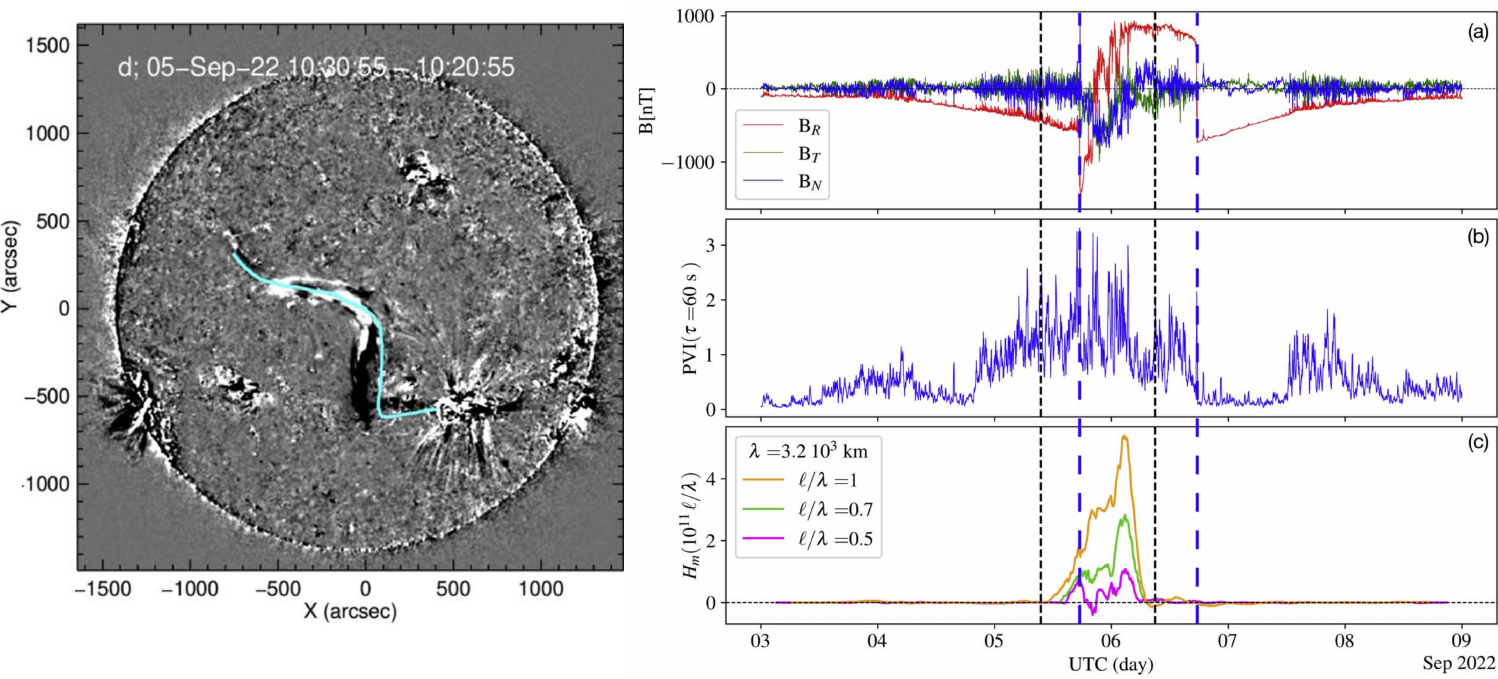} 
\caption{The SO observations and PSP measurements of a filament eruption on 5 September 2022. Left panel: A running difference image using SO EUI/FSI 174~\AA\ to show the plasma flows along the filament prior to eruption, indicating the presence of a flux rope structure. The line indicates the filament channel. Right panel: The corresponding flux rope detected in situ by PSP. The plots show (from top to bottom) the magnetic field measured by PSP/FIELDS, the PVI signal and the magnetic helicity at different length scales. The dashed lines represent the start and end of the flux rope structure whereas, the longdashed lines indicate discontinuities magnetic field. Figure adapted from \citet{long2023}.
} \label{fig:SO_PSP_fr}
\end{center}
\end{figure}

Another filament eruption occurred on 5 September 2022 on the farside of the Sun, where both PSP and SO were positioned at the time, allowing for the determination of the magnetic flux rope structure before its arrival \citep{long2023}. The eruption occurred just prior to SO's second perihelion, during its nominal mission phase, while undertaking a gravity assist maneuver at Venus. The erupting filament was observed by the remote-sensing instruments on board SO that was 150$^{\circ}$ ahead of Earth, at a distance of 0.72~au (left panel of Figure~\ref{fig:SO_PSP_fr}). The corresponding ICME was detected in situ at PSP on 6 September 2022 (right panel of Figure~\ref{fig:SO_PSP_fr}), while it was undergoing its 13th encounter and was at a close distance of 0.062~au, making it the closest ICME ever detected to date.

The filament was observed to be rooted in AR 13066 and extended across a large portion of the quiet Sun as seen by EUI/FSI. Prior to eruption, two similar plasma flows were observed to propagate from the active region across the quiet Sun. These plasma flows exhibited a helical right-handed motion suggesting the presence of a flux rope with at least one turn and positive helicity. By analysing the SPICE synoptic raster scans of the region, it was determined that the composition of the filament plasma within the flux rope was photospheric, suggesting that the flux rope formed as a result of magnetic reconnection via photospheric flux cancellation.

The filament erupted on 5 September 2022 and its ICME counterpart was detected in situ by PSP a day later and also by SO \citep{davies2024, trotta2024}. Although the flux rope had a complicated structure, it was identified in the magnetic field data taken by the FIELDS instrument on PSP by applying the magnetic helicity–partial variance of increments
(H$_{m}$-PVI) technique described in \citet{pecora2021}. By using this technique, it was possible to identify the helical structure of the ICME between 5-6 September 2022. The H$_{m}$-PVI technique determined that the flux rope detected has a positive helicity, which matches the handedness of the flux rope determined from EUI. This study highlights the importance of having both remote-sensing observations and in situ measurements at close distances and far from the Sun-Earth line but also long term observations of pre-eruptive structures and their source regions. 

\begin{figure}[t]
\begin{center}
 \includegraphics[width=5in]{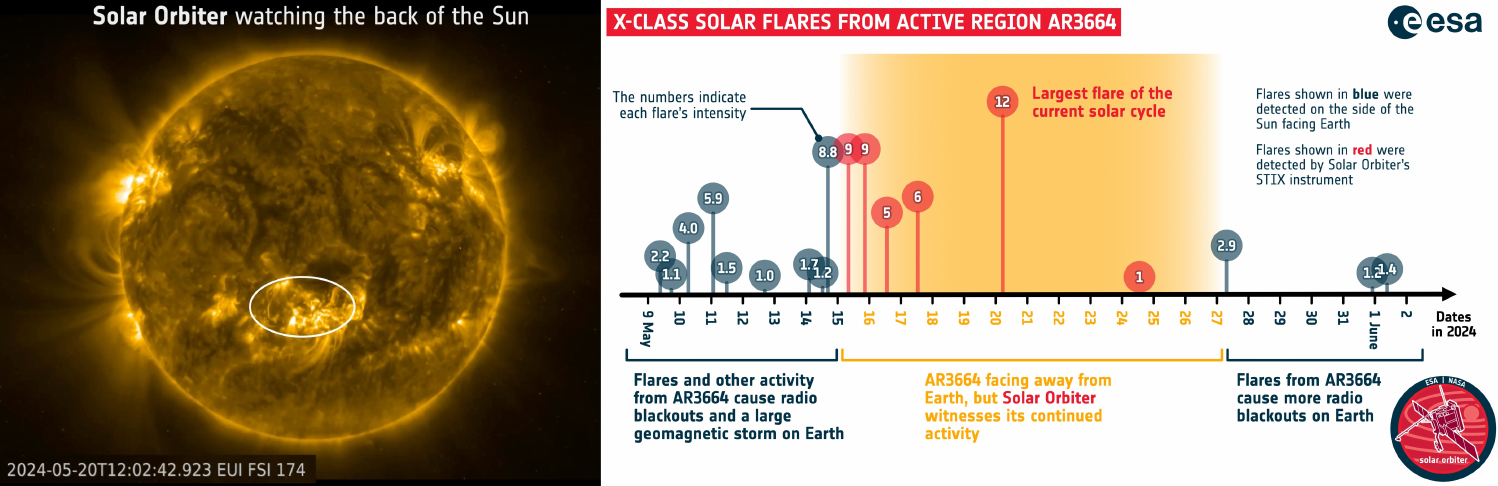} 
\caption{The observations and activity measured by Solar Orbiter during May 2024. Left panel: SO/EUI FSI 174~\AA\ image of AR 13664 as seen on the farside of the Sun. Right panel: Timeline of X-class flares from AR 13664 during its first solar rotation. Image courtesy of ESA \url{https://www.esa.int/Science_Exploration/Space_Science/Solar_Orbiter/Can_t_stop_won_t_stop_Solar_Orbiter_shows_the_Sun_raging_on}
} \label{fig:SO_may2024}
\end{center}
\end{figure}

As we head into the maximum of Solar Cycle 25, solar activity and the occurrence of geomagnetic storms are increasing along with the likelihood of PSP and SO observing these storms. In May 2024, an enormous active region (AR 13664) appeared on the Sun, surviving for 3 whole rotations. During its first rotation, it produced a multitude of strong X class flares and along with CMEs in quick succession, leading to the occurrence of the third largest storm in over 20 years and the sixth largest geomagnetic storm since 1957 \citep{hayakawa2024}, and also resulting in a solar energetic particle event that led to a ground-level enhancement. The AR rotated off the Earth-facing side of the Sun on the 14 May 2024 where it became visible to SO that was on the farside (Figure~\ref{fig:SO_may2024}). This allowed SO along with PSP and other heliospheric spacecraft to monitor the continued high level of activity from this region as it rotated across the farside of the Sun. In fact, it produced the largest flare of the solar cycle at the time, which STIX detected with an estimated class of X12 on 20 May 2024, which was immediately followed by an SEP event detected by EPD. This flare was associated with a CME which was seen by Metis and one day later was detected in situ by MAG. The ability to monitor and observe this active region on the farside mean that we had advanced warning (4-5 days) of what level of solar activity and hence resulting space weather effects we might experience before the region rotated back into Earth view. This provides an insight into how useful a space weather monitor positioned at L5, such as the upcoming Vigil mission, or at other vantage points (e.g. L4), in being able to monitor active region evolution and activity.

\section{Concluding Remarks}
This work highlights the importance of and the unique scientific opportunities enabled by Solar Orbiter and Parker Solar Probe and supporting space-borne assets and ground-based telescopes. Huge coordination efforts between the instrument teams and the international scientific community have already yielded exciting results in order to understand the origins, release and acceleration of the solar wind and also the structure, eruption and evolution of coronal mass ejections. There are still a wealth of opportunities to analyse SO-PSP datasets from both past and upcoming perihelia in order to maximise the scientific returns from these missions. The final three perihelion encounters from the nominal mission phase of PSP beginning December 2024 and also the increase in inclination of SO ($>$17$^{\circ}$ in February 2025) in order to observe the poles are imminent. The observations and measurements taken by PSP and SO at multiple vantage points (i.e. close heliocentric distances, farside of the Sun) have emphasized the importance of a whole realistic view of the Sun, which will hopefully be provided in the near future by upcoming missions such as Vigil at L5 \citep{palomba2022}, an L4 mission \citep{cho2023}, and Firefly: the 4$\pi$ constellation mission \citep{raouafi2023b}.

\acknowledgements
S.L.Y. is grateful to the Science Technology and Facilities Council for the award of an Ernest Rutherford Fellowship (ST/X003787/1). 
Parker Solar Probe was designed, built, and is now operated by the Johns Hopkins Applied Physics Laboratory as part of NASA’s Living with a Star (LWS) program (contract NNN06AA01C). We would like to acknowledge the efforts of the FIELDS, SWEAP, and ISOIS instrument and science operations teams and the PSP spacecraft engineering and operations team at the Johns Hopkins Applied Physics Laboratory. 
Solar Orbiter is a space mission of international collaboration between ESA and NASA, operated by ESA. The SO/EUI instrument was built by CSL, IAS, MPS, MSSL/UCL, PMOD/WRC, ROB, LCF/IO with funding from the Belgian Federal Science Policy Office (BELSPO/PRODEX PEA 4000134088); the Centre National d’Etudes Spatiales (CNES); the UK Space Agency (UKSA); the Bundesministerium f\"{u}r Wirtschaft und Energie (BMWi) through the Deutsches Zentrum f\"{u}r Luft- und Raumfahrt (DLR); and the Swiss Space Office (SSO). The German contribution to SO/PHI is funded by the BMWi through DLR and by MPG central funds. The Spanish contribution is funded by FEDER/AEI/MCIU (RTI2018-096886-C5), a “Center of Excellence Severo Ochoa” award to IAA-CSIC (SEV-2017-0709), and a Ramón y Cajal fellowship awarded to DOS. The French contribution is funded by CNES. The development of SPICE has been funded by ESA member states and ESA. It was built and is operated by a multi-national consortium of research institutes supported by their respective funding agencies: STFC RAL (UKSA, hardware lead), IAS (CNES, operations lead), GSFC (NASA), MPS (DLR), PMOD/WRC (Swiss Space Office), SwRI (NASA), UiO (Norwegian Space Agency). Solar Orbiter Solar Wind Analyser (SWA) data are derived from scientific sensors which have been designed and created, and are operated under funding provided in numerous contracts from the UK Space Agency (UKSA), the UK Science and Technology Facilities Council (STFC), the Agenzia Spaziale Italiana (ASI), the Centre National d’Etudes Spatiales (CNES, France), the Centre National de la Recherche Scientifique (CNRS, France), the Czech contribution to the ESA PRODEX programme and NASA. Solar Orbiter SWA work at UCL/MSSL is currently funded under STFC grants ST/W001004/1 and ST/X/002152/1. Solar Orbiter magnetometer operations are funded by the UK Space Agency (grant ST/T001062/1). Funding for SwRI was provided by NASA contract NNG10EK25C. Funding for the University of Michigan was provided through SwRI subcontract A99201MO. 
For the purpose of open access, the author has applied a Creative Commons Attribution (CC BY) licence (where permitted by UKRI) to any Author Accepted Manuscript version arising.


\end{document}